\documentclass{webofc}
\usepackage[varg]{txfonts}   
\usepackage{tikz}
\usetikzlibrary{positioning}
\usepackage{floatrow}

\begin{document}
\title{Search for the Chiral Magnetic and Vortical Effects Using Event Shape Approaches
in Au+Au Collisions at STAR}
%
%

\author{\firstname{Zhiwan} \lastname{Xu} (for the STAR Collaboration)\inst{1}\fnsep\thanks{\email{zhiwanxu@physics.ucla.edu} }
}

\institute{Department of Physics and Astronomy, University of California, Los Angeles, California 90095, USA}

\abstract{%
The chiral magnetic/vortical effect (CME/CVE)
in heavy-ion collisions probes the topological sector of Quantum Chromodynamics, where $\cal P$ and $\cal CP$ symmetries are violated locally in strong interactions. 
However, the experimental observables for the CME/CVE are dominated by backgrounds related to elliptic flow and nonflow. 
We employ event shape approaches to mitigate the flow background and event planes based on spectators to minimize the nonflow background.
We report on the CME search in Au+Au collisions at $\sqrt{s_{\rm NN}}$ = 7.7, 14.6, 19.6, 27, and 200 GeV, as well as the CVE search at 19.6 and 27 GeV.
}
\maketitle
\section{Introduction}
\label{intro}
The chiral magnetic effect (CME)~\cite{CME-1} originates from the interplay between an intense magnetic field ($\vec{B}$) and the chirality imbalance ($\mu_5$) of quarks,
which induces the electric charge separation, $\vec{J_{e}} \propto \mu_{5}\vec{B}$, in heavy-ion collisions. 
Similarly, a large vorticity field ($\vec{\omega}$) from the global angular momentum creates a baryonic charge separation, 
$\vec{J_{B}} \propto \mu_{5}\mu_{B}\vec{\omega}$, known as the chiral vortical effect (CVE)~\cite{CVE}. 
Both the CME and the CVE violate local ${\cal P}$ and $\cal{CP}$ symmetries in strong interactions and have stimulated extensive searches in heavy-ion collisions over the past two decades.
 
We employ the widely-used three-point correlator~\cite{CME-9}, $\gamma^{112} = \langle \cos(\varphi_{1}+\varphi_2-2\Psi_{\rm{RP}})\rangle$,
to detect charge separation, where $\varphi_i$ and $\Psi_{\rm{RP}}$ are the azimuthal angles of a final-state particle and the reaction plane, respectively.
The CME/CVE contributes positively to the difference between opposite-sign and same-sign pairs, $\Delta\gamma^{112} = \gamma^{\rm{OS}}- \gamma^{\rm{SS}}$.
The major background in $\Delta\gamma^{112}$ arises from the collective motion of particles, characterized by elliptic flow, $v_{2}=\langle \cos 2(\varphi-\Psi_{\rm{RP}})\rangle$.
Elliptic flow coupled with effects such as resonance decay, local charge conservation (LCC), and transverse momentum conservation, gives 
a positive contribution to $\Delta\gamma^{112}$.
In this work, we adopt event shape variables to categorize collision events and project $\Delta\gamma^{112}$ to the zero-flow intercept.
In addition, the nonflow contribution is minimized by using the event plane based on spectators, $\Psi_1$, as a proxy of $\Psi_{\rm{RP}}$, from STAR Event Plane Detector (EPD) and Zero Degree Calorimeter Shower Max Detector (ZDC-SMD).
Moreover, the spectator plane is closely correlated with the magnetic field direction.


\section{Method}

Schematically, event shapes are influenced by both the initial eccentricity and the final-state emission pattern. 
To primarily control eccentricity, the Event Shape Engineering (ESE)~\cite{SergeiESE,STAR-ESE} method was proposed to construct the event shape variable based on a sub-event B that excludes particles of interest (POI), $q_{2,\rm{B}} = \sqrt{\Bigl[(\sum_{i=1}^{N}{\sin 2\varphi_i })^2+(\sum_{i=1}^{N}{\cos 2\varphi_i})^2\Bigr]/N}$.
However, even at zero $q_{2,\rm{B}}$, the $v_2$ value for POI is still sizable, 
leading to an extrapolation over a large range to achieve $v_2
\rightarrow0$ limit introducing substantial fitting uncertainties.

In reality, the experimental data could be dominated by the event-by-event fluctuation in the emission pattern. 
Thus, a novel Event Shape Selection (ESS)\cite{ESS} method was developed to access both the initial geometry and the emission pattern.
We directly use POI to construct event shape variables, $q^2_2\{{\rm{POI}}\} = \Bigl[\bigl(\sum^N_{i=1} \sin2\varphi_i\bigr)^2 + \bigl(\sum^N_{i=1} \cos2\varphi_i\bigr)^2\Bigr]/\bigl[N(1+N\langle v_2\rangle ^2)\bigr].$
The normalization is improved with the next leading term of $N\langle v_2\rangle ^2$.
To further suppress the residual background due to the intrinsic correlation between $q^2_2$ and $v_2$, both of which are built from POI, particle pair information is introduced by adding the momenta of the two single particles, $\varphi_{\rm{P}} =\arctan \frac{p_{1,y} + p_{2,y}}{p_{1,x} + p_{2,x}}$. This scheme also better mimics the decay and LCC mechanisms.
Consequently, both $q_2$ and $v_2$ could be constructed from  $\varphi_{\rm{P}}$.
For better clarity, we add the label ``single" to those variables constructed using single particle momenta

Altogether, four ESS recipes have been established: (a) single $q^2_{2,\rm{POI}}$ - single $v_2$,
(b) pair $q^2_{2,\rm{POI}}$ - pair $v_2$,
(c) pair $q^2_{2,\rm{POI}}$ - single $v_2$, and
(d) single $q^2_{2,\rm{POI}}$ - pair $v_2$.
Extensive model studies using the event-by-event anomalous-viscous fluid dynamics (EBE-AVFD)~\cite{AVFD} show that
the optimal ESS approach is recipe (c), classifying events according to pair $q_2^2$ and projecting $\Delta\gamma^{112}$ to zero single $v_2$.

\section{Results}

\begin{figure}[tbhp]
\vspace{-0.3cm}
\begin{minipage}[c]{\textwidth}
\includegraphics[width=\textwidth]{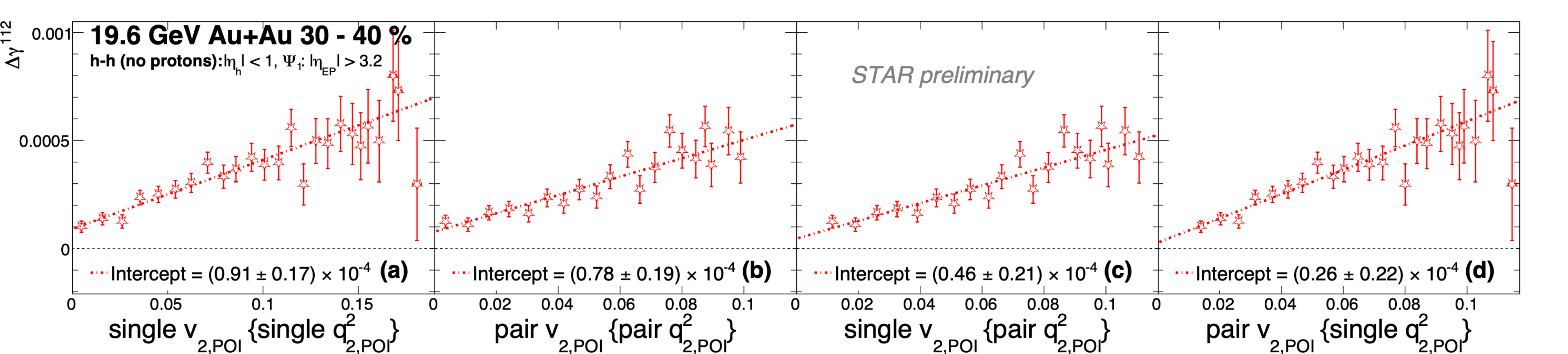}
\end{minipage}
\caption{$\Delta\gamma^{112}$ vs $v_2$ for hadrons or hadron pairs (excluding $p$ and $\Bar{p}$) with $|\eta|<1$, using spectator plane from EPD ($|\eta|>3.2$) in the 30--40\% centrality range of Au+Au collisions at 19.6 GeV.
$q_2^2$ and $v_2$ are based on either single particles or particle pairs from POI.
}
\vspace{-0.2cm}
\label{fig-1}            
\end{figure}

The application of four combinations of shape observables in ESS to STAR data of Au+Au collisions at 19.6 GeV is demonstrated in Fig.~\ref{fig-1} in the 30--40\% centrality range.
POI are hadrons within $|\eta|<1$ and $0.2<p_{T}<2$ GeV/$c$, excluding $p$ and $\Bar{p}$. 
(Anti-)Protons are rejected to suppress transported quark effects at lower beam energies.
The spectator plane ($\Psi_1$) is constructed with EPD hits at $\eta_{\rm{EPD}} > y_{\rm{beam}}$. 
For instance, $y_{\rm{beam}}=3.04$ at 19.6 GeV, so we apply $\eta_{\rm{EPD}}>3.2$ to enrich spectator contributions.
With each ESS recipe, a clear linear relation between $\Delta\gamma^{112}$ and single/pair $v_2$ is observed towards the zero-$v_2$ region. 
We retrieve $\Delta\gamma^{112}_{\rm{ESS}} = {\rm{Intercept}}\times(1-\langle v_2\rangle)^2$ based on the extrapolated intercept~\cite{ESS}.
\begin{figure}[ht]
\begin{minipage}[c]{0.45\textwidth}
\includegraphics[width=\textwidth]{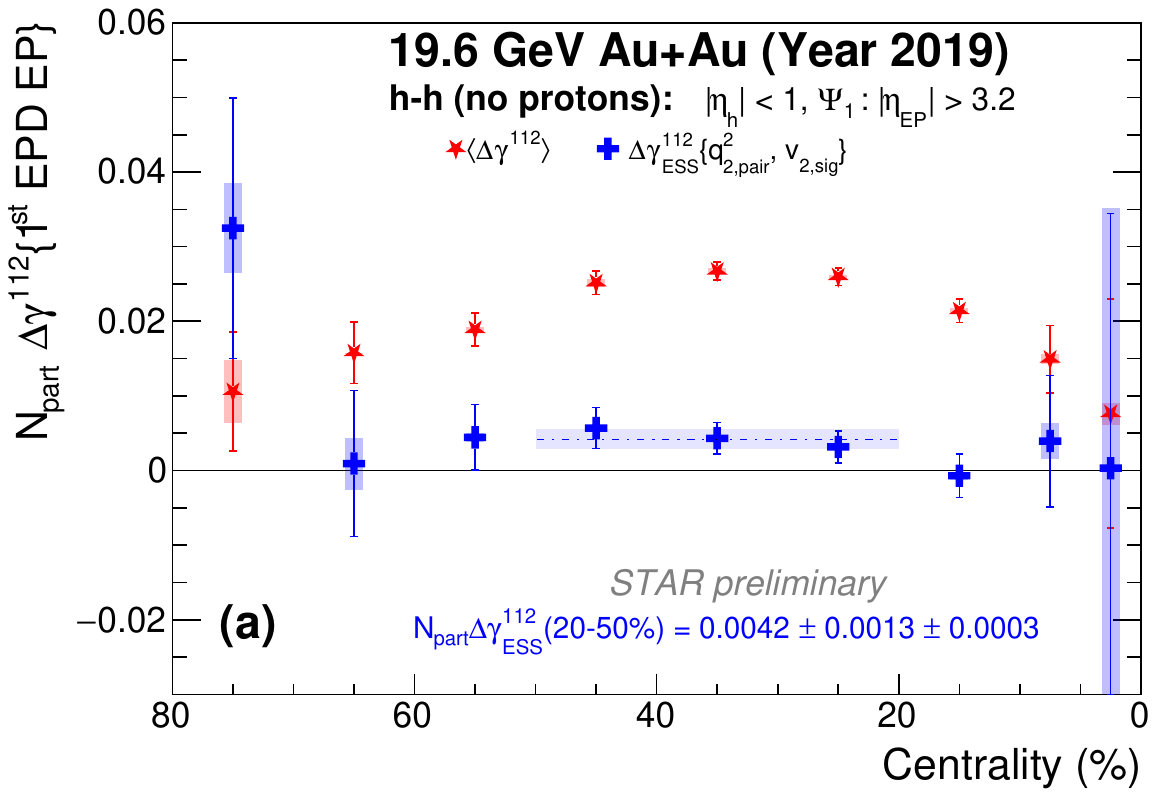}
\end{minipage}\hspace{0.2cm} 
\begin{minipage}[c]{0.45\textwidth}
\includegraphics[width=\textwidth]{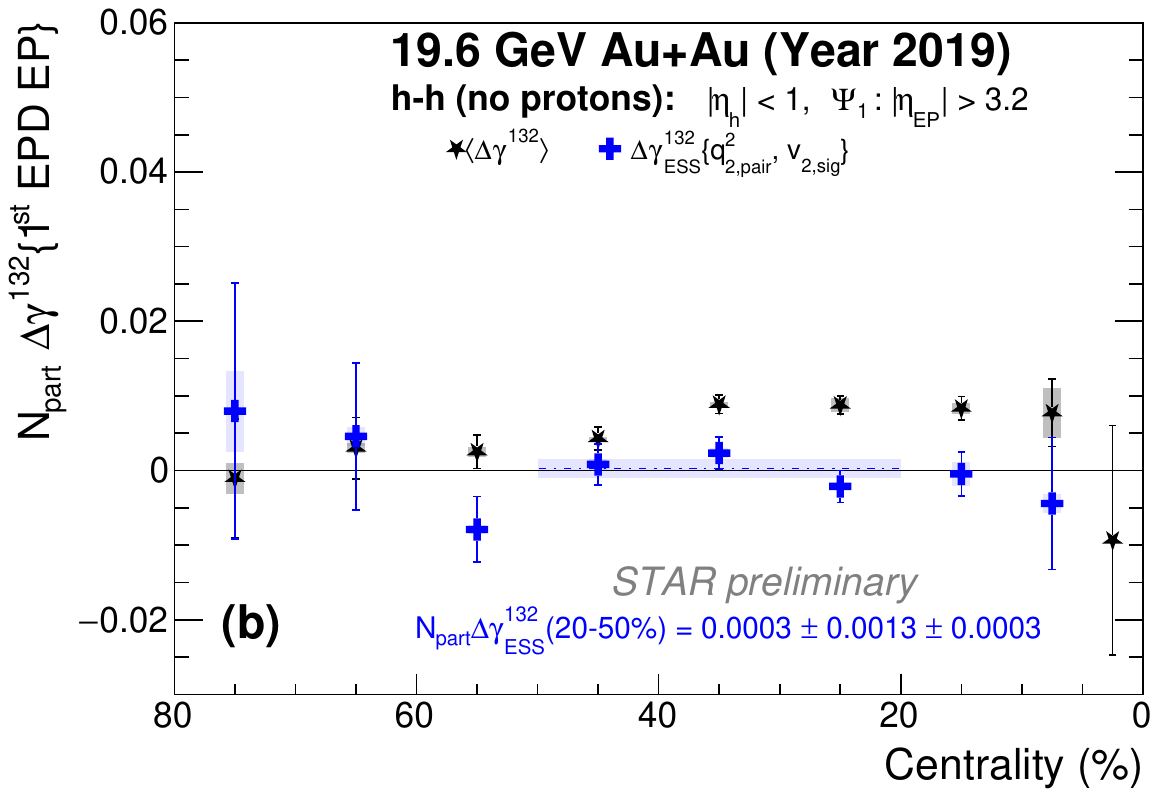}
\end{minipage} 
\caption{
(a) Centrality dependence of $N_{\rm{{part}}}~\Delta\gamma^{112}$ before and after the optimal ESS is applied in Au+Au at 19.6 GeV.  (b) The corresponding results for the background indicator, $N_{\rm{{part}}}~\Delta\gamma^{132}$. }
\label{fig-cme} 
\vspace{-0.3cm}
\end{figure}

\begin{figure}[h!]
\begin{minipage}[c]{0.70\textwidth}
    \includegraphics[width=1\textwidth]{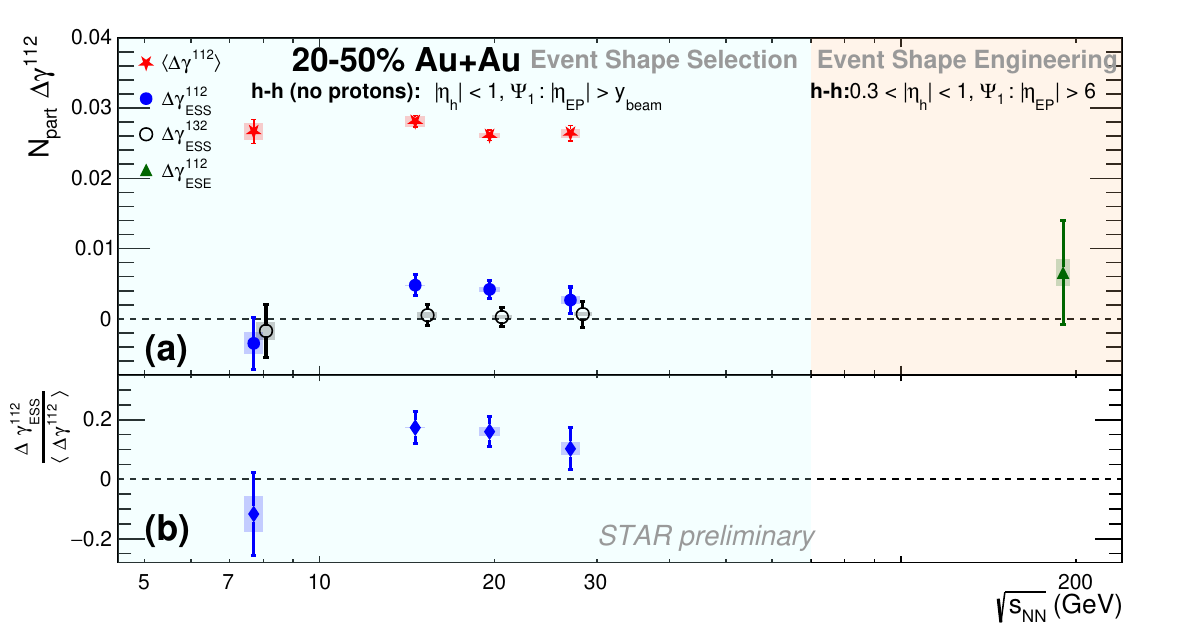}
\end{minipage} 
\begin{minipage}[c]{0.225\textwidth}
    \includegraphics[width=1\textwidth]{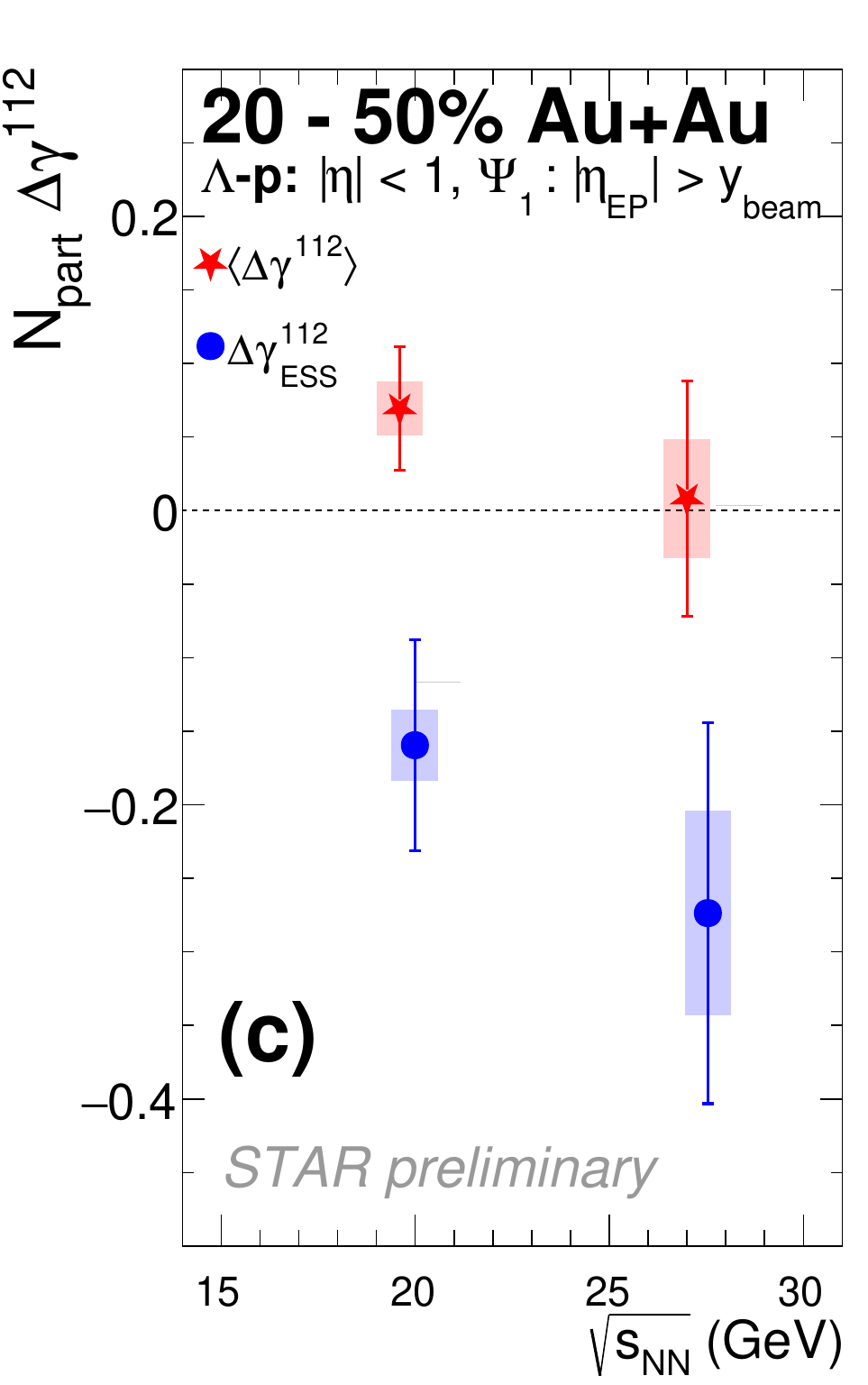}
\end{minipage} 
        
\caption{(a) Beam energy dependence of $N_{\rm{part}}\langle \Delta\gamma^{112}\rangle$,  $N_{\rm{part}}\Delta\gamma^{112}_{\rm{ESS}}$, and $N_{\rm{part}}\Delta\gamma^{132}_{\rm{ESS}}$ for $h$-$h$ (no $p$ and $\Bar{p}$) using spectator plane from EPD in Au+Au at 7.7, 14.6, 19.6, and 27 GeV. 
At 200 GeV, $N_{\rm{part}}\Delta\gamma^{112}_{\rm{ESE}}$ is presented for $h$-$h$ using spectator plane from ZDC-SMD. 
(b) Ratio of $\Delta\gamma^{112}_{\rm{ESS}}$ to the ensemble average.
(c) $N_{\rm{part}}\langle\Delta\gamma^{112}\rangle$ and $N_{\rm{part}}\Delta\gamma^{112}_{\rm{ESS}}$ for $\Lambda$-$p$ using spectator plane from EPD in Au+Au at 19.6 and 27 GeV.
}
\label{fig:cme-results}
\vspace{-0.2cm}
\end{figure}

Figure~\ref{fig-cme}(a) presents the centrality dependence of the ensemble average  $N_{\rm{{part}}}\langle\Delta\gamma^{112}\rangle$, as well as $N_{\rm{{part}}} \Delta\gamma^{112}_{\rm ESS}$ using the optimal ESS (c) in 0--80\% Au+Au at 19.6 GeV. $N_{\rm{{part}}}$ denotes the number of participating nucleons.
A constant fit over the 20--50\% centrality range yields a finite  $N_{\rm{{part}}}\Delta\gamma^{112}_{{\rm{ESS}}}$ value, with a $3\sigma$ significance.
Fig.~\ref{fig-cme}(b) shows the corresponding results for the background indicator, $\gamma^{132} = \langle \cos(\varphi_{1}-3\varphi_2+2\Psi_{{\rm{RP}}})\rangle$, which is known to be dominated by background.
After extrapolation to zero flow,
a constant fit over the 20--50\% centrality range renders a $N_{\rm{{part}}}\Delta\gamma^{132}_{{\rm{ESS}}}$ value consistent with zero. We have performed the same ESS analysis on the STAR data of Au+Au collisions at 7.7, 14.6, and 27 GeV, and the results will be summarized in Fig.~\ref{fig:cme-results}.


\begin{figure}
\vspace{-0.2cm}
\floatbox[{\capbeside\thisfloatsetup{capbesideposition={right,top},capbesidewidth=0.32\textwidth}}]{figure}[\FBwidth]
{\caption{(a) $\Delta\gamma^{112}$ vs single $v_2$ for all hadrons ($0.3<|\eta|<1$), using $\phi_c$ as a proxy of reaction plane from the sub-event other than that for the two POI, in 20--50\% Au+Au at 200 GeV. The event shape variable is $q_{2,{\rm{B}}}$.
(b) Intercepts obtained with different invariant mass windows.}\label{fig-2}}
{
\begin{minipage}[c]{0.3\textwidth}
\includegraphics[width=\textwidth]{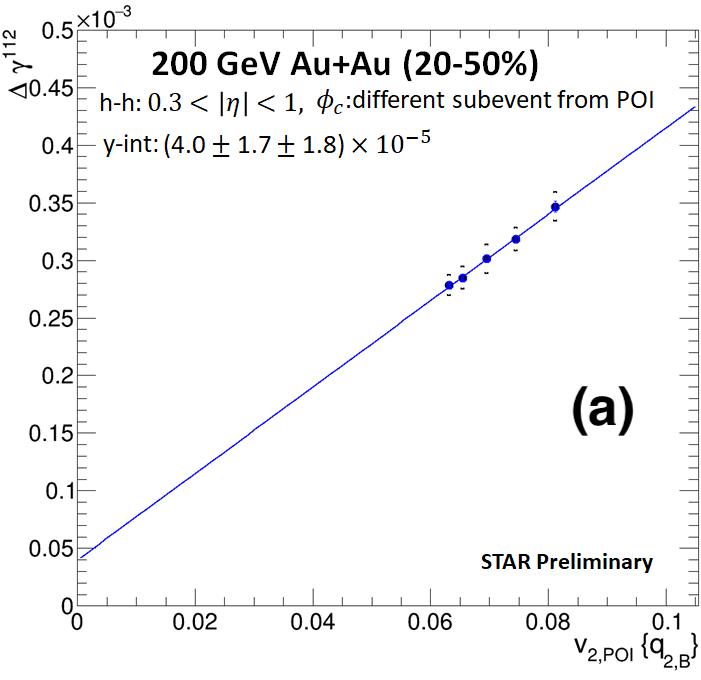}
\end{minipage}\hspace{0.2cm}
\begin{minipage}[c]{0.29\textwidth}
\includegraphics[width=\textwidth]{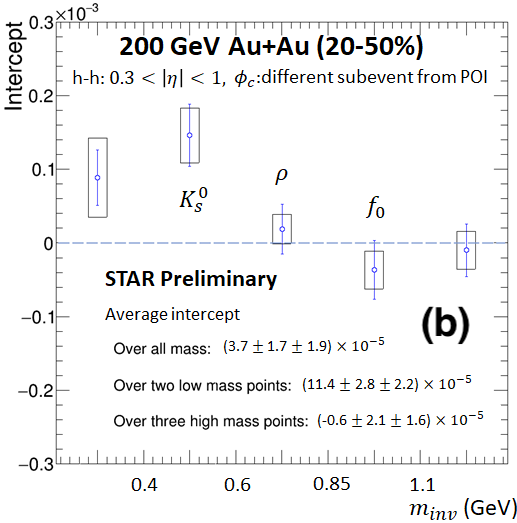}
\end{minipage}
}
\vspace{-0.4cm}
\end{figure}

Figure~\ref{fig-2}(a) shows results from another analysis with the application of previously proposed ESE
to Au+Au data at 200 GeV in 10\%-size centrality bins and averaged over the centrality range of 20--50\%.
The $q_{2,\rm{B}}$ is constructed with non-POI hadrons from sub-event B at midrapidities ($|\eta|<0.3$). 
In each event class categorized by $q_{2,\rm{B}}$,
the POI for $\Delta\gamma^{112}$ and $v_{2}$ measurements and the particles that serve as the event plane come from $0.3<\eta<1$ and $-1<\eta<-0.3$, respectively, or the other way around.
To project to the CME-sensitive intercept $\Delta\gamma^{112}_{{\rm{ESE}}}$, a long extrapolation is needed.

Furthermore, we have investigated various invariant mass windows in the Au+Au data at 200 GeV, which offers more insights into the background. 
The invariant mass of a particle pair is constructed by adding their four momenta, assuming that both particles are pions. 
A non-zero $\Delta\gamma^{112}_{{\rm{ESE}}}$ is observed in the low-mass region ($M_{inv}<0.6$ GeV/$c^2$). In this analysis, the POI and the reference particle for the event plane are all from midrapidities, so nonflow effects~\cite{Nonflow} are present and need to be addressed. To mitigate nonflow, the ESE analysis over the entire invariant mass region is performed with the first-order event plane from the ZDC-SMD. The intercept is consistent with zero with a large uncertainty.

Figure~\ref{fig:cme-results}(a) reports the beam energy dependence of the hadron-hadron $\Delta\gamma^{112}$ results in 20--50\% Au+Au collisions. The ESS and ESE measurements utilize the EPD and ZDC-SMD event planes, respectively, minimizing nonflow effects.
At both 14.6 and 19.6 GeV, the $N_{\rm{part}}\Delta\gamma^{112}_{\rm{ESS}}$ values are positive with a 3$\sigma$ significance.
At 7.7 GeV, $N_{\rm{part}}\Delta\gamma^{112}_{\rm{ESS}}$ is consistent with zero within the current uncertainty.
At 27 and 200 GeV, the statistical uncertainties are large to make definitive conclusions.
Meanwhile, $N_{\rm{part}}\Delta\gamma^{132}_{\rm{ESS}}$ is consistent with zero at all available beam energies.

Figure~\ref{fig:cme-results}(b) shows the ratio of $\Delta\gamma^{112}_{\rm{ESS}}/\langle \Delta\gamma^{112}\rangle$, which illustrates a significant reduction of flow and nonflow backgrounds in our new ESS approach.
We have identified that at least 80\% of the $\Delta\gamma^{112}$ constituents arise from the background. 
The disappearance of the ESS observable at 7.7 GeV 
is consistent with the disappearance of CME signal  
if the partonic degrees of freedom or chiral symmetry restoration disappears
at such temperature and $\mu_{B}$.

Figure~\ref{fig:cme-results}(c) presents the $\Lambda$-$p$ $\Delta\gamma^{112}$ measurements in search of the CVE  using spectator plane from EPD in 20--50\% Au+Au collisions at 19.6 and 27 GeV.
$\Lambda$ and ${\bar \Lambda}$ with $|\eta|<1$ and $0.2<p_{T}<2$ GeV/$c$ are reconstructed using decay daughters of pions and (anti)protons.
The (anti)protons that enter the $\Delta\gamma^{112}$ measurements have excluded the decay daughters of $\Lambda$. 
After background subtraction,  $N_{\rm{part}}\Delta\gamma^{112}_{\rm{ESS}}$ in 20–50\% centrality range renders negative values,  
therefore the signature of CVE signals remains inconclusive at either of the energies.

\section{Summary}
We have exploited event shape variables to suppress the flow background in $\Delta \gamma^{112}$ to search for the CME and the CVE in Au+Au collisions at RHIC.
We employ a broad spectrum of observables and analysis techniques including particle pair information, different invariant mass windows, and spectator planes from EPD or ZDC-SMD.
After categorizing events based on their shapes and extrapolating the CME/CVE observable to the zero-flow intercept, we report the $\Delta \gamma^{112}$ measurements using $h$-$h$  correlations in Au+Au at $\sqrt{s_{\rm NN}}$ = 7.7, 14.6, 19.6, 27, and 200 GeV, and those using $\Lambda$-$p$ at 19.6 and 27 GeV.
We observe a finite $\Delta \gamma^{112}_{{\rm{ESS}}}$ intercept value with a 3$\sigma$ significance in Au+Au at 14.6 and 19.6 GeV, whereas background  
measure $\Delta \gamma^{132}_{{\rm{ESS}}}$ is consistent with zero.
A finite $\Delta \gamma^{112}_{{\rm{ESE}}}$ intercept is also found at the low-mass region in Au+Au at 200 GeV, where nonflow effects need to be addressed in order to relate the low mass $\Delta \gamma^{112}_{{\rm{ESE}}}$ intercept to the CME.

\end{document}